\documentclass[aps,prl,reprint,twocolumn]{revtex4-1}
\usepackage{graphicx}
\usepackage{amsmath}
\usepackage{amssymb}
\usepackage[colorlinks,allcolors=blue]{hyperref}
\usepackage{hypcap}

\newcommand{\pref}[2]{\hyperref[#1]{\ref{#1}#2}}
\newcommand{\eqpref}[1]{\hyperref[#1]{(\ref{#1})}}

\begin{document}
\title{Interacting atomic quantum fluids on momentum-space lattices}
\author{Bryce~Gadway}
\email{bgadway@illinois.edu}
\author{Fangzhao~Alex~An}
\author{Eric~J.~Meier}
\author{Jackson~Ang'ong'a}
\affiliation{Department of Physics, University of Illinois at Urbana-Champaign, Urbana, IL 61801-3080, USA}
\date{\today}
\begin{abstract}
We study the influence of atomic interactions on quantum simulations in momentum-space lattices (MSLs), where driven atomic transitions between discrete momentum states mimic transport between sites of a synthetic lattice. Low energy atomic collisions, which are short ranged in real space, relate to nearly infinite-ranged interactions in momentum space. However, the distinguishability of the discrete momentum states coupled in MSLs gives rise to an added exchange energy between condensate atoms in different momentum orders, relating to an effectively attractive, finite-ranged interaction in momentum space. We explore the types of phenomena that can result from this interaction, including the formation of chiral self-bound states in topological MSLs. We also discuss the prospects for creating squeezed states in momentum-space double wells.
\end{abstract}
\pacs{PACS1 ; PACS2}
\maketitle

Quantum simulation with ultracold atoms~\cite{Bloch-RMP08,AtomsRev-NatPhys-2012} has been a powerful tool in the study of many-body physics and nonequilibrium dynamics. There has been recent interest in extending quantum simulation studies from real-space potentials to synthetic lattice systems composed of discrete internal~\cite{Boada-Synth,Celi-ArtificialDim} or external~\cite{NateGold-TrapShake} states. These synthetic dimensions enable many unique capabilities for quantum simulation, including new approaches to engineering nontrivial topology~\cite{Celi-ArtificialDim,Wall-SpinOrb}, access to higher dimensions~\cite{Boada-Synth}, and potential insensitivity to finite motional temperature.

The recent development of momentum-space lattices (MSLs), based on the use of discrete momentum states as effective sites, has introduced a fully synthetic approach to simulating lattice dynamics~\cite{Gadway-KSPACE,Meier-AtomOptics,Meier-SSH,Alex-2Dchiral,Alex-Annealed}. As compared to partially synthetic systems~\cite{Fallani-chiral-2015,Stuhl-Edge-2015}, fully synthetic lattices offer complete microscopic control of system parameters. While this level of control is analogous to that found in photonic simulators~\cite{SzameitReview-2010,PhotRev-NatPhys-2012}, matter waves of atoms can interact strongly with one another.

However, fully synthetic systems also present apparent challenges for studying nontrivial atomic interactions. Synthetic systems based purely on internal states suffer from limited state spaces, sensitivity to external noise for generic, field-sensitive states~\cite{Sugawa-NonAb}, and possible collisional relaxation~\cite{Soding-relaxation} and three-body losses~\cite{Weiner-Collisions}. Furthermore, for isotropic scattering lengths as in $^{87}$Rb~\cite{Sugawa-NonAb} and alkaline earth atoms~\cite{Pagano-AlkalineEarth}, interactions in the synthetic dimension are nearly all-to-all. Similarly, $s$-wave contact interactions relate to nearly infinite-ranged momentum-space interactions at low energy, and should naively be decoupled from particle dynamics in MSLs.

Here, we investigate the role of atomic interactions in MSLs, showing that finite-ranged interactions in momentum space result from the exchange energy of bosonic condensate atoms in distinguishable momentum states. We explore potential interaction-driven phenomena that can be studied in topological MSLs, showing that chiral propagating bound states can emerge in the presence of an artificial magnetic flux. We additionally discuss the use of momentum-space double wells for the generation of squeezed many-particle states.

MSLs provide a bottom-up approach to engineering designer Hamiltonians with field-driven transitions. This technique is based on the coherent coupling of multiple atomic momentum states via two-photon Bragg transitions, synthesizing an effective lattice of coupled modes in momentum space. In the general case, the transition frequency associated with each Bragg transition is unique. For free non-interacting particles, this stems from the quadratic dispersion $E_p^0 = p^2/2m$, with momentum $p$ and atomic mass $m$. Considering atoms initially at rest and driven by laser fields of wavelength $\lambda$ and wavevector $k = 2\pi/\lambda$, a discrete set of momentum states $p_n = 2n\hbar k$ may be coupled, having energies $4 n^2 E_r$, with $E_r = \hbar^2 k^2 / 2m$ being the photon recoil energy. By individually addressing the unique Bragg transition resonances, one may realize MSLs with full ``local'' and temporal parameter control. Specifically, single-particle tight-binding models of the form
\begin{equation}
H^{sp} \approx \sum_n t_n(e^{i \varphi_n} \hat{c}^\dag_{n+1} \hat{c}_n + \mathrm{h.c.}) + \sum_n \varepsilon_n \hat{c}^\dag_n \hat{c}_n \ ,
\label{EQ:e00}
\end{equation}
may be realized by a single pair of Bragg lasers, where $\hat{c}_n$ ($\hat{c}^\dagger_n$) is the annihilation (creation) operator for the state with momentum $p_n$. Here, nearest-neighbor tunneling elements are controlled through the amplitude and phase of individual frequency components of the Bragg laser field, which drive first-order, two-photon Bragg transitions~\cite{Kozuma-Bragg}. Similarly, an effective potential landscape of site energies $\varepsilon_n$ is controlled by small frequency detunings of the laser fields from Bragg resonances.

\begin{figure}[t!]
\includegraphics[width=\columnwidth]{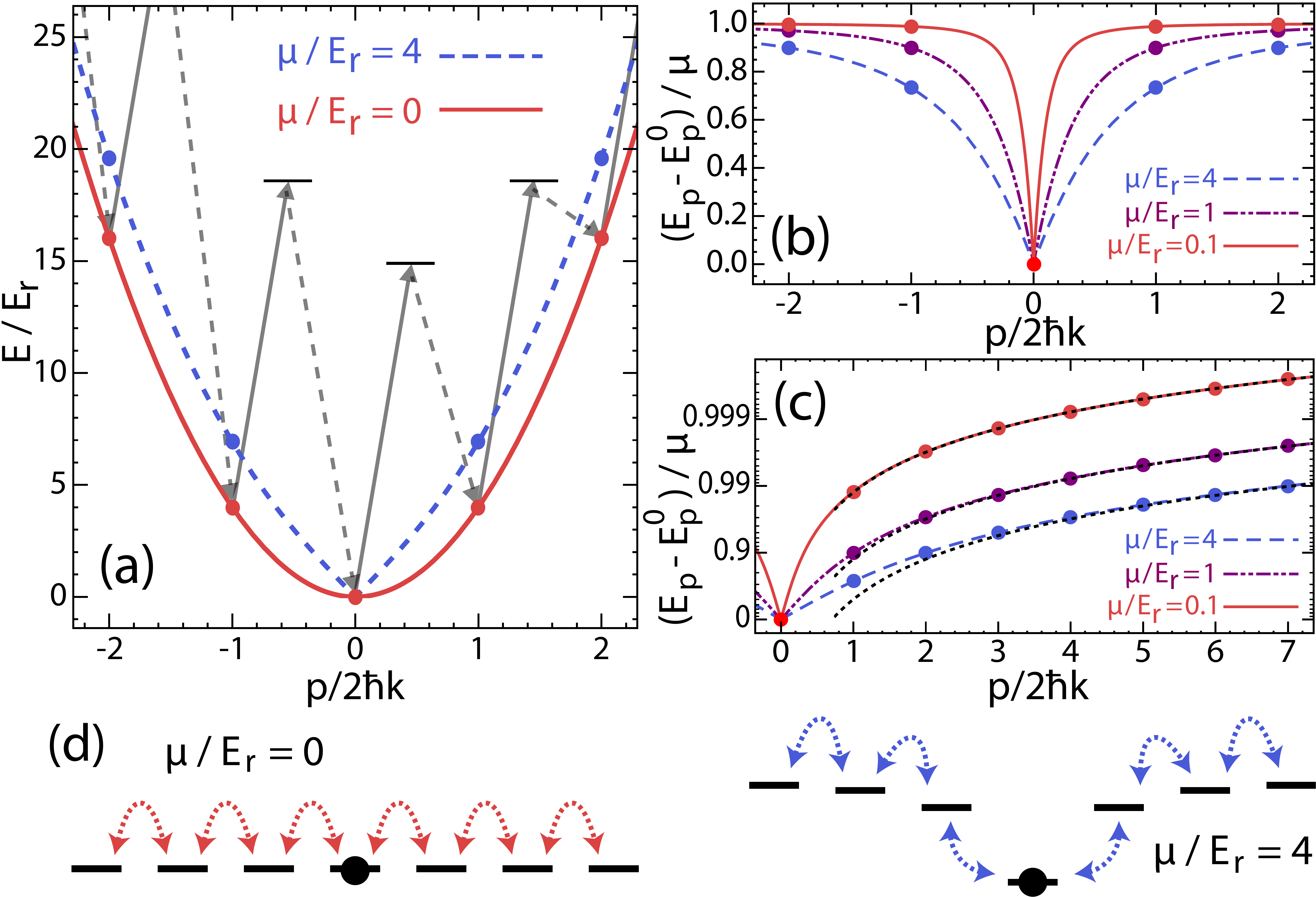}
\caption{\label{FIG:fig1}
\textbf{Interaction shifts of Bragg tunneling resonances.}
\textbf{(a)}~Energy dispersion of non-interacting massive atoms $E_p^0$ (red solid line) in units of the recoil energy $E_r$, and the Bogoliubov dispersion $E_p$ of a homogeneous gas with weak repulsive interactions and a mean-field energy $\mu = 4 E_r$ (blue dashed line).
\textbf{(b)}~The effective interaction potential  (normalized to $\mu$) experienced by weakly-coupled excitations with momentum $p$, shown for the cases $\mu / E_r = \ $0.1, 1, and 4.
\textbf{(c)}~Semi-log plot of the effective interaction potentials in (b), shown for a larger range of momenta, compared to the form $\mu - \mu^2/2 E_p^0$ (dotted lines) relevant in the free-particle limit ($E_p^0 \gg 2\mu$).
\textbf{(d)}~Cartoon depiction of effective site energies shifted by central population density, for $\mu/E_r = 0$ (no interactions) and $\mu/E_r = 4$.
}
\end{figure}

While there have been several demonstrations~\cite{Meier-AtomOptics,Meier-SSH,Alex-2Dchiral,Alex-Annealed} of the ability to engineer diverse single-particle Hamiltonians using MSLs, the prospects for studying interactions and correlated dynamics have not yet been examined. In typical real-space atomic quantum simulations, two-body contact interactions are the dominant mechanism leading to correlated behavior~\cite{Greiner-SF-MI-2002}. The two-body contact potential $V(\textbf{r},\textbf{r'})$, being nearly zero ranged in real space, relates to a nearly infinite-ranged interaction potential $V(\textbf{k},\textbf{k'})$ in momentum space. At first glance, it appears as though only all-to-all interactions should result (considering only mode-preserving interactions), which unfortunately cannot give rise to correlated behavior for a fixed total density.

However, we find that finite-ranged, attractive interactions arise in the MSL due to atom statistics in the quantum fluid. As the landscape of MSL site energies is determined by synthesized detunings from the Bragg transition resonances, an interaction potential results from density-dependent modifications to the free-particle energy dispersion. We consider the case of small-amplitude momentum excitations of a homogeneous bosonic quantum gas at rest with uniform particle density $n$. The quadratic dispersion $E_p^0$ for a non-interacting gas is shown in Fig.~\pref{FIG:fig1}{(a)}, along with that of Bogoliubov quasiparticles of a weakly-interacting quantum gas~\cite{Stenger-Bragg,Ozeri-Bog-RMP,Hadzibabic-Bragg-Strong}, $E_p = \sqrt{E_p^0 (E_p^0 + 2\mu)}$ (ignoring a uniform energy shift of $\mu$ for all states). Here, $\mu = g n$ is the uniform condensate mean-field energy, with the interaction parameter $g$ related to the $s$-wave scattering length $a$ as $g = 4\pi \hbar^2 a/m$.

\begin{figure*}[t!]
\includegraphics[width=\textwidth]{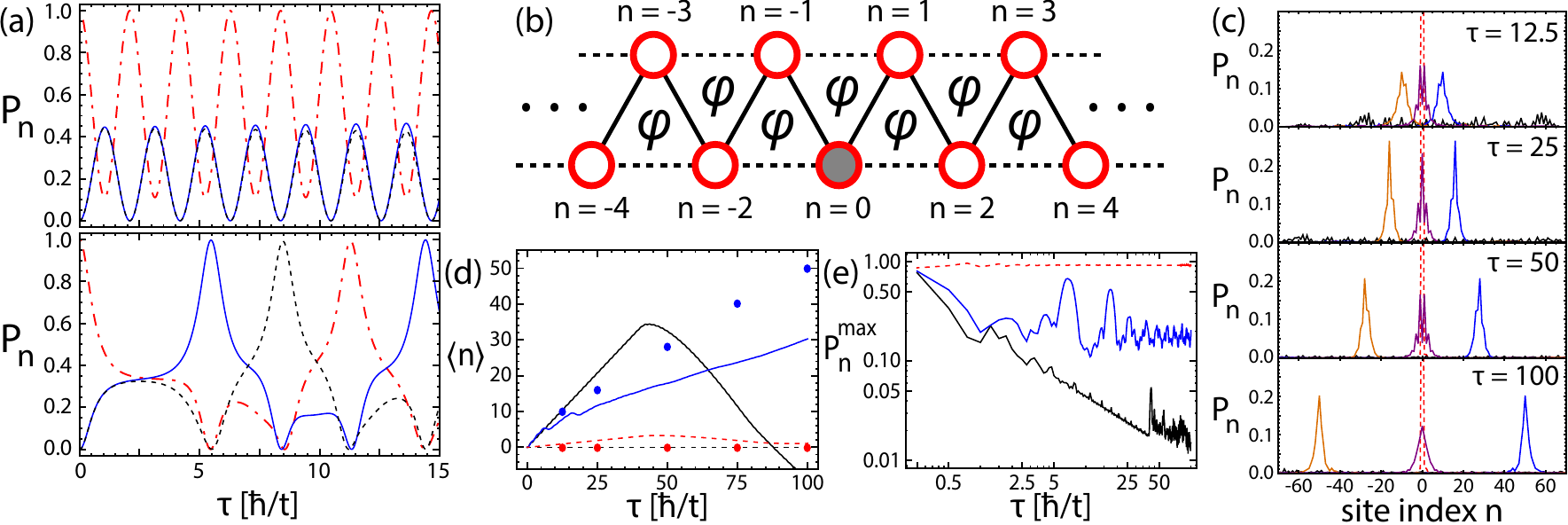}
\caption{\label{FIG:fig2}
\textbf{Interaction effects in multiply-connected momentum-space lattices.}
(\textbf{a})~Role of interactions on quench dynamics of atoms in a momentum-space triple well with periodic boundary conditions and an effective magnetic flux of $\varphi = 0.001 \pi$. Population begins in site $n=0$ (red dashed-dotted line), and equal magnitude tunnelings $t$ to sites $n = 1$ (blue solid line) and $n = -1$ (black dashed line) are turned on at time $\tau = 0$. The upper plot shows the dynamics of the site populations $P_n$ for $\mu/t = 0$ (no interactions), and the lower plot for the case $\mu / t = 6$.
(\textbf{b})~Cartoon depiction of atoms on a zig-zag ladder with uniform magnetic flux $\varphi$ and population initialized at the central site $n=0$ (gray). Tunnelings realized by first- (solid black lines) and second-order (dashed black lines) Bragg transitions have uniform amplitudes.
(\textbf{c})~Snapshots of the distributions of site populations for increasing evolution times $\tau = 12.5, 25, 50,$ and 100 (units of $\hbar / t$), shown for several different combinations of interaction-to-tunneling ratios ($\mu/t$) and flux values ($\varphi$). Solid black denotes $\{\mu/t , \varphi \} = \{0 , \pi/6\}$, solid blue for $\{7.2 , \pi/6\}$, dashed red for $\{12 , \pi/6\}$, solid purple for $\{7.2 , 0\}$, and solid orange for $\{7.2 , -\pi/6\}$.
(\textbf{d})~Dynamics of the average site position $\langle n \rangle$ of the atomic distributions, shown for the cases of $\{\mu/t , \varphi \} = \{0 , \pi/6\}$ (black solid line), $\{7.2 , \pi/6\}$ (blue solid line), and $\{12 , \pi/6\}$ (red dashed line). The blue and red dots relate to the position of largest site population for the cases $\mu/t = 7.2$ and 12, respectively. The reflection of the black line near $\tau = 40 \hbar/t$ results from population reaching the boundary of the 401-site system.
(\textbf{e})~Dynamics of $P_n^{max}$, the percentage of atoms in the most highly-populated site, for the same cases as in (d). The spike of the black line near $\tau = 40 \hbar/t$ is also due to boundary reflection in the non-interacting case.
}
\end{figure*}

The interactions mainly modify the shape of the free-particle dispersion at low momenta, with a characteristic linear dispersion near $p = 0$. At higher momenta ($E_p^0 \gg 2\mu$), the Bogoliubov quasiparticles are free-particle-like and have a roughly quadratic dispersion which is shifted in energy by the chemical potential ($E_p \approx E_p^0 + \mu - \mu^2 / 2 E_p^0$). This extra energy shift of order $\mu$ for high momentum states is a consequence of exchange interactions with the zero-momentum condensate. Indeed, the modification of the energy-momentum dispersion results not from momentum-dependent collisional interactions, but rather from the quantum statistics of the identical bosons in distinguishable motional states~\cite{exch,Kaufman-Entang,Larson-MultibandBosons}. This is analogous to the effective magnetic interactions of electrons in condensed matter that result from spin-independent Coulomb interactions and exchange statistics.

The shifts of the state energies from their non-interacting values ($E_p - E_p^0$) are plotted in Fig.~\pref{FIG:fig1}{(b,c)}. The Bragg transition frequencies, which depend only on the energy differences between pairs of adjacent momentum states, are most strongly perturbed near the initially populated zero-momentum site. The density-dependent perturbations to the transition frequencies result in an effective interaction potential that is finite ranged and attractive, as depicted in Fig.~\pref{FIG:fig1}{(d)}.
For $\mu \gtrsim E_r$, the momentum-space interaction potential has significant off-site contributions, the effective range of which increases with larger ratios $\mu / E_r$. There is, however, a natural limitation on the compatibility of long-ranged interactions with the scheme for engineering MSLs. This method breaks down when unique spectral addressing of the individual Bragg transitions is lost, occurring when multiple momentum orders populate the linear phonon branch (occurring roughly when $\mu$ exceeds $8 E_r$, the bare energy spacing of the Bragg resonances).

For concreteness, we now focus on the limit $\mu \ll 2 E_r$, where all coupled momentum orders are approximately distinguishable and the interaction potential is effectively local in momentum space. This allows us to move beyond a description of weakly-coupled condensate excitations, and describe the more general case where atomic population is arbitrarily distributed among many momentum orders. We explore new phenomena that may be opened up to investigation by combining this simple local interaction with the wide range of tunable lattice models enabled by MSLs.

The role of momentum-space interactions in the limit of purely distinguishable momentum orders can be simply described by a multimode nonlinear Schr\"{o}dinger equation, where the self- and cross-phase modulation terms describing intra-mode and inter-mode interactions differ by the exchange energy. We make the simplifying assumption that all momentum orders share a common spatial wavefunction throughout the dynamics. This single-mode approximation is valid on only relatively short timescales, and does not capture the spatial separation of momentum wavepackets. To focus on the unique contributions of this exchange-driven momentum-space interaction, we additionally assume single-spin (internal state) bosonic atoms with effectively one-dimensional dynamics. In this restricted scenario, four-wave mixing processes~\cite{Deng-FWM-1999,Trippenbach-FWM-theory,Rolston-NL-2002,Pertot-10-PRL} are not allowed, and the individual state populations are conserved by the atomic interactions.

In this single-mode approximation, assuming a homogeneous condensate with fixed total atom number $N$ and thus fixed density, we may represent the condensate wavefunction simply with appropriately normalized (to unity) complex amplitudes $\phi_n$ of the various discrete plane-wave momentum orders with momenta $p_n = 2 n \hbar k$~\cite{Trippenbach-FWM-theory}. We furthermore remove a global energy term $2\mu$ by redefining the $\phi_n$, thus transforming the momentum-space interaction into an effectively attractive self-interaction term for atoms residing in the same order (valid for $\mu \ll 2E_R$).
Taking into account the contributions of these interactions to the effective tight-binding models of Eq.~\ref{EQ:e00}, the dynamical evolution of the atoms becomes governed by
\begin{equation}
i \hbar \dot{\phi}_n = \sum_m H^{sp}_{mn} \phi_m - \mu |\phi_n|^2 \phi_n \ ,
\label{EQ:e0d}
\end{equation}
a tunable lattice tight-binding model with local attractive interactions.

Even with only two coupled sites, interactions are expected to significantly alter the system dynamics. For the simple case of population initialized in one of two equal-energy sites, weak interactions ($\mu/t < 4$) lead to a slowdown of two-mode Rabi dynamics, giving way to critical slowing for $\mu / t \approx 4$. For stronger interactions ($\mu/t > 4$), nonlinear self trapping~\cite{Smerzi-Joseph-Atoms-1997,Raghavan-Jos-1999} occurs, preventing population from ever fully leaving the initially-populated site. Moreover, many analogs of behavior found in tunnel-coupled superconductors, including plasma oscillations, the ac and dc Josephson effects, hysteresis, and macroscopic quantum self trapping, can be expected to emerge in double wells of Bragg-coupled states~\cite{Smerzi-Joseph-Atoms-1997,Raghavan-Jos-1999}.

While such interaction-driven phenomena are well studied in the two-mode case~\cite{Albiez-Oberth-2005,Levy-ACDC-2007,Leblanc-Joseph,Eckel-Hysteresis-2014,Trenkwalder-ParSymmBreak-2016,Chang-spinor-josephson,Tomko-Oberth-2017}, MSLs offer unique capabilities for engineering multiply-connected lattice geometries~\cite{Ryu-SQUID}. In particular, we consider the effects that local, attractive interactions can have on particle dynamics in MSLs with closed tunneling pathways, where artificial magnetic fluxes play a nontrivial role~\cite{Alex-2Dchiral,An-prep}. In Fig.~\ref{FIG:fig2}, we explore atom dynamics on lattices with triangular and rhomboidal geometries, attainable in one physical dimension with a combination of first- and second-order Bragg transitions (nearest- and next-nearest-neighbor tunnelings, respectively)~\cite{An-prep}.
We first consider the simplest such configuration, consisting of three sites with periodic boundary conditions, which is an analog of multiply-connected superconducting quantum interference devices or atomtronic circuits~\cite{Ryu-SQUID}.

Figure~\pref{FIG:fig2}{(a)} illustrates triple-well dynamics in the presence of a very weak applied flux ($0.001 \pi$) and uniform tunneling amplitude $t$. Without interactions (upper plot), an initially localized wavepacket spreads almost evenly to the neighboring sites. For sufficiently large interactions ($\mu/t = 6$, lower plot), however, the initial onset of a slightly asymmetric chiral current induces the formation of a fully chiral soliton-like mode~\cite{Tromb-Breather}. For still larger values of the nonlinear interaction, self trapping occurs.

We extend this investigation to a many-site zig-zag ladder system, shown in Fig.~\pref{FIG:fig2}{(b)}, with a uniform distribution of effective magnetic fluxes $\varphi$ and tunnelings $t$. The role of nonlinear interactions in such a topological lattice model is of interest for its connection to emergent topological phenomena in kinetically frustrated systems. Here we again examine the case of population initialized to a single, central mode ($n=0$), exploring dynamics following a tunneling quench. The distributions of normalized site populations $P_n$ at various evolution times $\tau$ are shown in Fig.~\pref{FIG:fig2}{(c)}. With no interactions (black solid line), chiral currents are present, but with a rapid ballistic spreading of the atomic distribution. We find that moderate interactions stabilize the atomic distribution, leading to soliton- or breather-like states~\cite{Tromb-Breather}. For non-zero flux values ($\pm \pi/6$ for the blue and orange solid lines), these self-stabilized states propagate in a chiral fashion along the zig-zag ladder. For much larger interaction strengths ($\mu/t \gtrsim 9$), the atoms remain localized at $n = 0$ for all flux values.

This interaction-driven stabilization of chiral wavepackets is further summarized in Fig.~\pref{FIG:fig2}{(d,e)}, through the average position $\langle n \rangle$ and largest site population $P_n^{max}$. Figure~\pref{FIG:fig2}{(d)} contrasts the dynamics of $\langle n \rangle$ for interaction values $\mu/t = 0, 7.2,$ and 12, all for a uniform magnetic flux $\varphi = \pi/6$. While some dynamics of $\langle n \rangle$ can be seen even for the self-trapped scenario at $\mu/t = 12$, the position of the most highly populated site (filled circles) never deviates from the initial location. The dynamics of this site population $P_n^{max}$ are shown in Fig.~\pref{FIG:fig2}{(e)} for these same cases. Without interactions, ballistic spreading leads to a continuous decrease of the maximum density. For a range of moderate interactions, we find that the distribution self stabilizes at short timescales ($\tau \approx \hbar/t$). Finally, for very large interactions, behavior analogous to macroscopic quantum self trapping inhibits particle transport and population remains largely localized at the central site.

\begin{figure}[t!]
\includegraphics[width=\columnwidth]{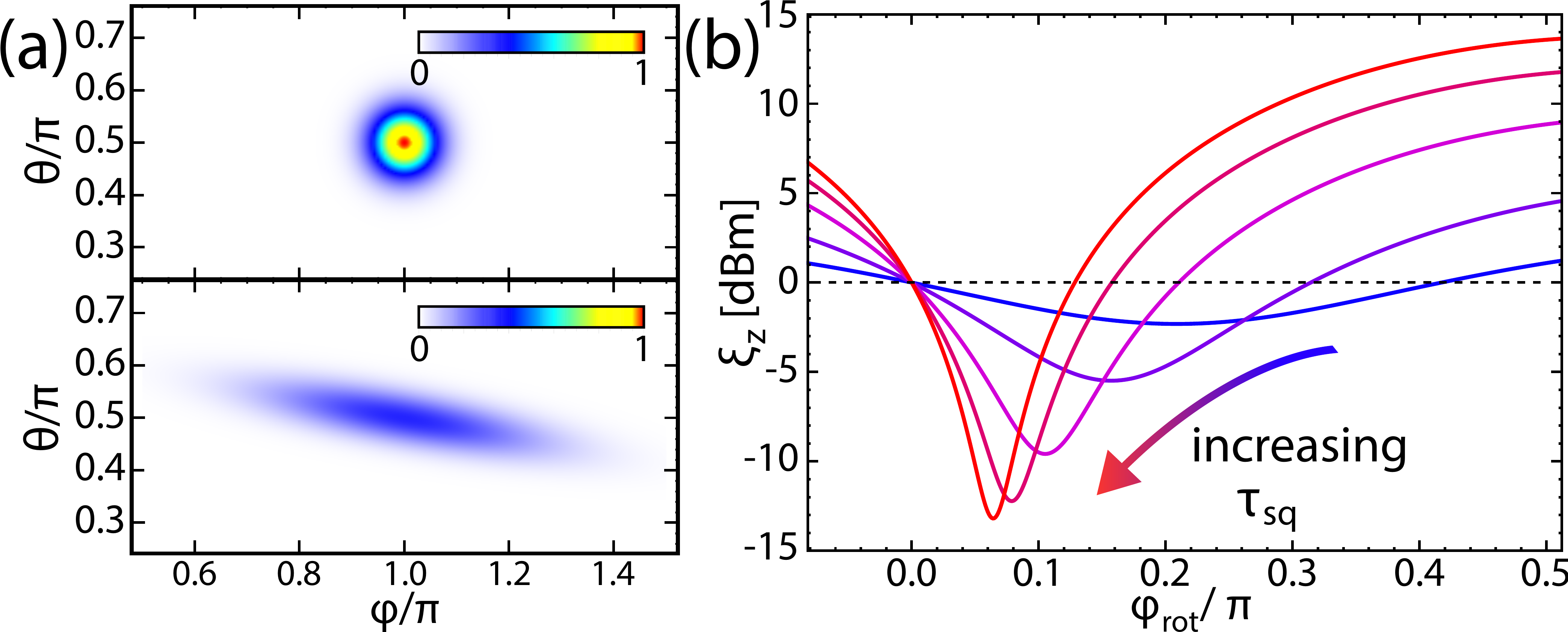}
\caption{\label{FIG:fig3}
\textbf{Squeezing in a momentum-space double well.}
(\textbf{a})~Visualization of many-particle ($N = 100$) spin states $|\Psi\rangle$ through their overlap with different coherent spin states $|\langle \theta,\varphi|\Psi\rangle|^2$. Shown are the cases of an initial coherent spin state $|\pi/2,\pi\rangle$ (upper plot), and the transformed state after evolution under $H_{\mathrm{sq}}$ for a time $\kappa \tau_{\mathrm{sq}}/\hbar = 0.0173 \pi$ (lower plot).
(\textbf{b})~Squeezing along the $\hat{z}$-axis, $\xi_z$, for different evolution times $\kappa \tau_{\mathrm{sq}}/\hbar = \{0.1,0.25,0.5,0.75,1\}\times 0.0173 \pi$ (solid lines, with colors varying from blue to red) and for different angles of rotation $\varphi_{\mathrm{rot}}$ of the final distribution about $\hat{J}_x$.
}
\end{figure}

Populating flat energy bands~\cite{DiamondLattice-2016,Vito-Flatband-14,Ehud-FlatBand-10,FlatBand-Nonlinear,Ani-synthz-zigzag} of similar topological models with interacting atoms should lead to interesting, emergent many-body dynamics. Interacting gases, combined with engineered MSLs having arbitrary and time-fluctuating disorder~\cite{Alex-Annealed}, should also enable highly controllable explorations into the physics of many-body localization~\cite{aleiner:finite_temperature_disorder_2010,Deissler-DisorderWithInteractions-2010}. For both scenarios, the most interesting open questions relate to phenomena driven by quantum fluctuations, which are not captured by Eq.~\ref{EQ:e0d}.

The simplest MSL to capture such physics is a single momentum-space double well. For fixed total particle number $N = N_1 + N_2$, one can define effective angular momentum operators relating to the coherences and macroscopic occupations $N_1$ and $N_2$ of two Bragg-coupled momentum orders (ignoring thermal and quantum depletion). These are given by $\hat{J}_x = (\hat{c}_1^\dag \hat{c}_2 + \hat{c}_2^\dag \hat{c}_1)/2$, $\hat{J}_y = i(\hat{c}_1^\dag \hat{c}_2 - \hat{c}_2^\dag \hat{c}_1)/2$, and $\hat{J}_z = (\hat{c}_2^\dag \hat{c}_2 - \hat{c}_1^\dag \hat{c}_1)/2$, where $\hat{c}_n$ ($\hat{c}^\dagger_n$) is the annihilation (creation) operator for mode $n$~\cite{RaghavanBigelow}. Here, Dicke states $|j,m\rangle$ with total spin $j = N/2$ and $\hat{z}$-projection $m = (N_2 - N_1) / 2$ describe collective two-mode number states. Coherent spin states (CSSs) of the form
$|\theta , \varphi\rangle = \sum_{m = -j}^{j} f_{m}^j (\theta, \varphi) e^{-i(j+m)\varphi}|j,m\rangle$, for
$f_{m}^j (\theta, \varphi) = \binom{2j}{j+m}^{1/2}\cos(\theta/2)^{j-m}\sin(\theta/2)^{j+m}$, result from a global rotation of spin-polarized states $|j,j\rangle$ about the spin vector $\hat{n}_\varphi = \cos(\varphi) \hat{J}_x + \sin(\varphi) \hat{J}_y$ by an amount $\theta$~\cite{CSS-RMP}. In this description, the momentum-space interaction relates to an effective nonlinear squeezing Hamiltonian $H_{\mathrm{sq}} = \kappa \hat{J}_z^2$, for $\kappa = -3\mu/2N$~\cite{RaghavanBigelow}.

We examine the case of the CSS $|\pi/2,\pi\rangle$ initially aligned along $-\hat{J}_x$. For short evolution times, the nonlinear Hamiltonian $H_{\mathrm{sq}}$ leads to a ``shearing'' of such coherent states. This is depicted in Fig.~\pref{FIG:fig3}{(a)}, for the initial CSS (upper plot) and the sheared, non-classical squeezed state after a time $\kappa\tau_{\mathrm{sq}}/\hbar = 0.0173 \pi$ (lower plot), through the overlap of these states with CSSs of varying $\theta$ and $\varphi$ values. For ease of calculation, dynamics are shown for the case of only $N = 100$ atoms ($j = 50$).

Figure~\pref{FIG:fig3}{(b)} shows, for sheared distributions relating to various 
evolution times $\tau_{\mathrm{sq}}$, the $\hat{z}$-axis squeezing parameter $\xi_z = 2 j \frac{\langle \Delta\hat{J}_z^2\rangle}{j^2 - \langle\hat{J}_z\rangle^2}$ as a function of rotation angle $\varphi_{rot}$ about the $\hat{J}_x$ spin axis. For typical experimental parameter values ($N = 10^5$ atoms, $\mu / \hbar = 2 \pi \times 1$~kHz), an optimal squeezing of $\xi_z^{min} \approx (3j^2)^{-1/3}$, relating to $-33$~dBm (after rotation), would be expected after a total duration $\tau_{\mathrm{sq}} \approx 1.13 j^{1/3}(\hbar/\mu) \approx 6.6$~ms~\cite{RaghavanBigelow}. The practical enhancement of atom interferometry from such squeezing will be challenged by decoherence and dephasing, such as from heating and laser phase noise. However, refocusing $\pi$ (echo) pulses or twist-and-turn squeezing schemes~\cite{Muessel-TWIST} can be used to mitigate decoherence due to spatial separation of the momentum orders. Looking beyond this simple two-mode case, the straightforward extension to multiple momentum states should also allow for unique investigations into multi-mode squeezing and quantum phase transitions in multi-mode analogs of the Lipkin-Meshkov-Glick model~\cite{Trenkwalder-ParSymmBreak-2016,Tomko-Oberth-2017}.

We have shown that local real-space interactions of atomic gases can give rise to finite-ranged effective interactions in momentum space.
These momentum-space interactions can lead to correlated dynamics of atoms in highly-tunable MSLs, opening up new possibilities for exploring interacting topological and disordered fluids. To focus on this effectively attractive momentum-space interaction, we have restricted our investigations to the case of mode-preserving collisions. It will be interesting to explore the effect of mode-changing collisions on MSL studies, with four-wave mixing relevant in higher dimensions~\cite{Deng-FWM-1999,Trippenbach-FWM-theory,Rolston-NL-2002} and for spinful condensates~\cite{Pertot-10-PRL}, as these give rise to effectively infinite-ranged, correlated hopping terms. While beyond the scope of this work, it will also be interesting to examine the roles of inhomogeneous density, finite temperature, elastic $s$-wave scattering, departures of the momentum excitations from the form of Bogoliubov quasiparticles~\cite{Hadzibabic-Bragg-Strong}, and energy-dependent modifications to the $s$-wave scattering length.


%

\end{document}